\documentclass[twocolumn,nofootinbib]{revtex4-2}
\usepackage{amsmath,amssymb,bm,graphicx,hyperref}
\allowdisplaybreaks[1]

\renewcommand\d{\partial}
\newcommand\<{\langle}
\renewcommand\>{\rangle}
\newcommand\eps{\varepsilon}
\renewcommand\k{{\bm{k}}}
\renewcommand\l{{\bm{l}}}
\newcommand\x{{\bm{x}}}
\newcommand\I{\mathcal{I}}
\newcommand\V{\mathcal{V}}
\newcommand\mm{\mathrm{mm}}
\newcommand\Mpc{\mathrm{Mpc}}
\DeclareMathOperator\sgn{sgn}

\begin{document}

\title{Circular polarization of the cosmic microwave background\\induced by the optical Magnus effect on gravitational lensing}

\author{Yusuke Nishida}
\affiliation{Department of Physics, Institute of Science Tokyo,
Ookayama, Meguro, Tokyo 152-8551, Japan}

\date{May 2026}

\begin{abstract}
Polarization of the cosmic microwave background (CMB) brings out information not only on the early universe but also on the late-time large-scale structure via weak gravitational lensing.
Here, we show that circular polarization is induced in principle from CMB temperature fluctuations when the optical Magnus effect is incorporated into gravitational lensing.
This is a consequence of the transverse shift of a trajectory of light depending on its helicity that requires right-handed and left-handed components at the same observation point to be sourced from different points of the surface of last scattering.
Whereas the resulting circular polarization is found far beyond the scope of current detection, our work establishes the optical Magnus effect on gravitational lensing as a new fundamental mechanism to produce circular polarization of CMB.
\end{abstract}

\maketitle

\section{Introduction}
Polarization of the cosmic microwave background (CMB) serves as one of the most sensitive probes of the early universe and the late-time large-scale structure~\cite{Planck:2020a}.
In the standard model of Big Bang cosmology, CMB polarization is produced by Thomson scattering of anisotropic radiation at the epoch of last scattering~\cite{Hu:1997,Cabella:2004}.
Because Thomson scattering preserves parity, it produces strictly linear polarization with circular polarization vanishing.
Therefore, any detection of circular polarization of CMB is expected to signal new astrophysical mechanisms or fundamental physics beyond the Standard Model of particle physics~\cite{Kamionkowski:1999,Samtleben:2007}.

Several mechanisms to produce circular polarization have been proposed.
For example, Faraday conversion of linear to circular polarization occurs when CMB photons propagate through magnetized relativistic plasmas, such as in galaxy clusters and supernova
remnants of the first stars~\cite{Cooray:2003,De:2015}.
Furthermore, circular polarization can be induced by Thomson scattering under primordial magnetic fields~\cite{Giovannini:2009,Giovannini:2010}, photon-photon scattering by Euler-Heisenberg interactions~\cite{Motie:2012}, and scattering of photons with the cosmic neutrino background~\cite{Mohammadi:2014} (see also Refs.~\cite{Montero-Camacho:2018,Hoseinpour:2020} for later clarifications).
From the perspective of fundamental physics, various extensions of the Standard Model are shown to predict circular polarization of CMB~\cite{Agarwal:2008,Finelli:2009,Alexander:2009,Zarei:2010}.
Prospects for its detection are closely evaluated in Ref.~\cite{King:2016}.

Whereas such mechanisms rely on specific foregrounds, early universe physics, or new fundamental interactions, weak gravitational lensing of CMB by the large-scale structure is an unavoidable secondary effect~\cite{Bartelmann:2001,Lewis:2006,Bartelmann:2010}.
With the intrinsic temperature and polarization fields remapped to observed ones, the conversion of linear polarization between $E$- and $B$-modes occurs, but circular polarization remains vanishing~\cite{Zaldarriaga:1998}.
The purpose of our work is to show that circular polarization of CMB is induced in principle from its temperature fluctuations when the optical Magnus effect is incorporated into gravitational lensing.

\section{Optical Magnus effect}
The optical Magnus effect arises as a correction to geometrical optics at the linear order in wavelength and causes transverse shift of a trajectory of light opposite for right-handed and left-handed circular polarizations~\cite{Dooghin:1992,Liberman:1992,Bliokh:2004a,Bliokh:2004b,Onoda:2004,Bliokh:2015,Ling:2017}.
It manifests itself not only when circularly polarized light propagates through an inhomogeneous optical medium but also when it propagates in a curved spacetime~\cite{Oancea:2019,Andersson:2023}.
Such a gravitational analog of the optical Magnus effect is sometimes referred to as the gravitational spin Hall effect of light and has been derived with various approaches~\cite{Berard:2006,Gosselin:2007,Frolov:2011,Yoo:2012,Yamamoto:2017,Yamamoto:2018,Duval:2017,Duval:2019,Oancea:2020,Frolov:2020,Andersson:2021,Harte:2022}.

Recently, the optical Magnus effect on gravitational lensing was studied in Ref.~\cite{Nishida:2026}, where the lens equation incorporating the optical Magnus effect was formulated.
In particular, under a weak gravitational potential $\phi=\phi(\x)\ll c^2$ in an expanding flat universe described by $ds^2=a(\eta)^2[-(1+2\phi/c^2)(cd\eta)^2+(1-2\phi/c^2)\delta_{ij}dx^idx^j]$, the modified lens equation in the first Born approximation reads $\bm\beta=\bm\theta-\bm\alpha-\lambda\tilde{\bm\alpha}$ with
\begin{align}
\alpha_i &= \frac2{c^2}\int_0^{\chi_*}\!d\chi\,\frac{\chi_*-\chi}{\chi_*\chi}\d_i\phi, \\
\label{eq:lens}
\tilde\alpha_i &= -\frac2{c^2k}\int_0^{\chi_*}\!d\chi\,\frac1{\chi_*\chi}\varepsilon^{ij}\d_j\phi.
\end{align}
Here, $\chi$ is the comoving radial distance from an observer at $\chi=0$ to a source at $\chi=\chi_*$, $\bm\beta$ and $\bm\theta$ are two-dimensional angular coordinates of source and image on the celestial sphere, respectively, $\phi$ in the integrands is evaluated at $\x=(-\chi\theta_1,\chi\theta_2,\chi)$, $\eps^{ij}$ is the Levi-Civita symbol, and $\d_i\equiv\d_{\theta_i}$ with $i=1,2$.
Whereas $\bm\alpha$ is the usual deflection angle by gravitational lensing~\cite{Bartelmann:2001,Lewis:2006,Bartelmann:2010}, it is $\lambda\tilde{\bm\alpha}$ that reflects the transverse shift of a trajectory of light due to the optical Magnus effect.
The latter is proportional to the comoving wavelength $2\pi/k$ as well as the helicity $\lambda=\pm1$ corresponding to right-handed and left-handed circular polarizations.
We note that the gravitational potential is assumed to vanish close to the observer to drop another term involving $\eps^{ij}\d_j\phi/\chi$ at $\chi=0$~\cite{Nishida:2026}.

\section{Circular polarization}
We now consider an intrinsic intensity of CMB denoted by $I(\bm\beta)=\sum_{\lambda=\pm1}I_\lambda(\bm\beta)$ with implicit dependence on $k$.
The vanishing circular polarization requires the equal intensity $I_\lambda(\bm\beta)=I(\bm\beta)/2$ for right-handed and left-handed components.
Because circularly polarized light observed at $\bm\theta$ is sourced from $\bm\beta=\bm\theta-\bm\alpha-\lambda\tilde{\bm\alpha}$, the lensed intensity is provided by $\I_\lambda(\bm\theta)=I_\lambda(\bm\theta-\bm\alpha-\lambda\tilde{\bm\alpha})$ for each component.
Therefore, right-handed and left-handed components at the same observation point actually correspond to different points of the intrinsic intensity.
Its fluctuations then lead to the usual lensed intensity,
\begin{align}
\I(\bm\theta) = \sum_{\lambda=\pm1}\I_\lambda(\bm\theta)
= I(\bm\theta) - \bm\alpha\cdot\bm\d I(\bm\theta) + O(\phi^2),
\end{align}
and furthermore,
\begin{align}\label{eq:circular}
\V(\bm\theta) = \sum_{\lambda=\pm1}\sgn(\lambda)\,\I_\lambda(\bm\theta)
= -\tilde{\bm\alpha}\cdot\bm\d I(\bm\theta) + O(\phi^2).
\end{align}
The resulting intensity difference between right-handed and left-handed components constitutes the circular polarization of CMB induced by the optical Magnus effect on gravitational lensing.

We introduce the lensing potential of the optical Magnus effect by $\tilde\alpha_i\equiv\varepsilon^{ij}\d_j\tilde\psi(\bm\theta)$ with Eq.~(\ref{eq:lens}).
The Fourier representation of the induced circular polarization in Eq.~(\ref{eq:circular}) reads
\begin{align}
\V(\l) = -\int\!\frac{d\l'}{(2\pi)^2}(\l\times\l')\,\tilde\psi(\l-\l')I(\l') + O(\phi^2)
\end{align}
in the flat-sky approximation.
Its angular power spectrum, defined by $\<\V(\l)\V(\l')\>\equiv(2\pi)^2\delta(\l+\l')C_{|\l|}^\V$ and analogously for the lensing potential and the intrinsic intensity, is provided by
\begin{align}\label{eq:polarization}
C_{|\l|}^\V = \int\!\frac{d\l'}{(2\pi)^2}(\l\times\l')^2C_{|\l-\l'|}^{\tilde\psi}C_{|\l'|}^I + O(\phi^3).
\end{align}
The power spectrum of the gravitational potential is defined by $\<\phi(\k)\phi(\k')\>\equiv(2\pi)^3\delta(\k+\k')P_\phi(|\k|)$ with $\phi(\k)$ being the Fourier representation of $\phi(\x)$.
It is related to the angular power spectrum of the lensing potential by
\begin{align}\label{eq:potential}
C_l^{\tilde\psi} = \frac1{(k\chi_*)^2}
\int_0^{\chi_*}\!d\chi\,\frac4{(c\chi)^4}P_\phi\!\left(\frac{l+\frac12}{\chi}\right)
\end{align}
in the Limber approximation~\cite{Bartelmann:2001,Lewis:2006,Bartelmann:2010,LoVerde:2008}.
Therefore, the angular power spectrum of the induced circular polarization can be evaluated with Eqs.~(\ref{eq:polarization}) and (\ref{eq:potential}) once the power spectrum of the gravitational potential and the angular power spectrum of the intrinsic intensity are provided.

\section{Angular power spectrum}
\begin{figure}[t]
\includegraphics[width=0.9\columnwidth,clip]{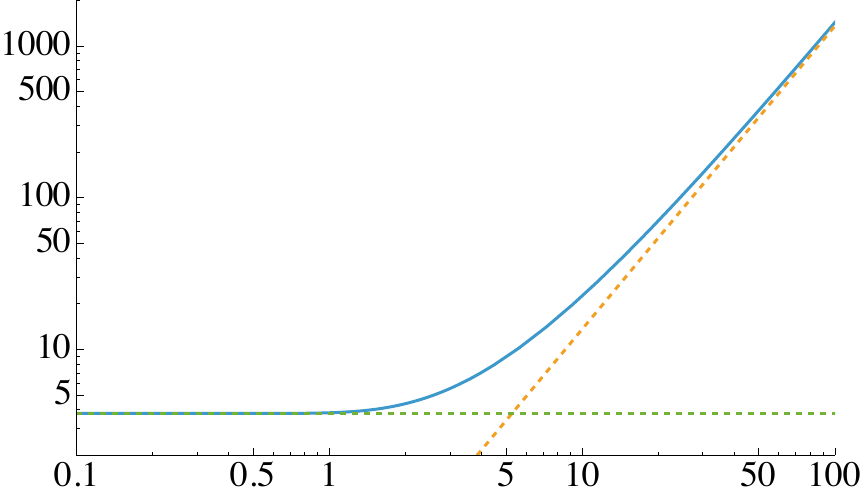}
\caption{\label{fig:prefactor}
$C_l^\V/(\bar{I}^2c_l^\V)\times10^{60}$ (vertical axis) in Eq.~(\ref{eq:spectrum}) as a function of $2\pi/k~[\mm]$ (horizontal axis).
The dashed lines indicate the asymptotic forms of $C_l^\V/(\bar{I}^2c_l^\V)\to1/(k\chi_*)^2$ at $w_k\ll1$ and $C_l^\V/(\bar{I}^2c_l^\V)\to[\hbar c/(\chi_*k_B\bar{T})]^2\approx3.8\times10^{-60}$ at $w_k\gg1$.}
\end{figure}

Planck's law relates the angular power spectrum of the intrinsic intensity to that of CMB temperature fluctuations by
\begin{align}
\frac{C_l^I}{\bar{I}^2} = \left(\frac{w_k}{1-e^{-w_k}}\right)^2\frac{C_l^T}{\bar{T}^2},
\end{align}
where $\bar{I}$ and $\bar{T}\approx2.725$ K are the mean intensity and temperature, respectively, and $w_k\equiv\hbar ck/(k_B\bar{T})\approx(5.28~\mm)\times k/(2\pi)$.
It is then convenient to express the angular power spectrum of the induced circular polarization as
\begin{align}\label{eq:spectrum}
C_l^\V \equiv \frac{\bar{I}^2}{(k\chi_*)^2}\left(\frac{w_k}{1-e^{-w_k}}\right)^2 \times c_l^\V,
\end{align}
so that $c_l^\V$ is dimensionless and independent of $k$.
With $\chi_*\approx14000~\Mpc$ adopted as the distance to the surface of last scattering, the prefactor of $c_l^\V$ is shown in Fig.~\ref{fig:prefactor} as a function of the wavelength of light.
For example, it is found as small as $C_l^\V/(\bar{I}^2c_l^\V)\approx22\times10^{-60}$ at $2\pi/k=10~\mm$ due to $k\chi_*\approx0.27\times10^{30}$.

\begin{figure}[t]
\includegraphics[width=0.9\columnwidth,clip]{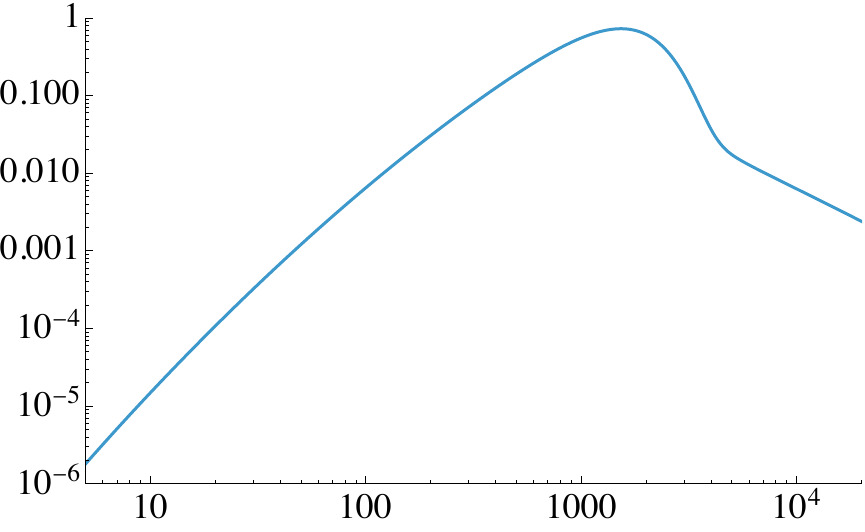}
\caption{\label{fig:spectrum}
$l(l+1)c_l^\V/(2\pi)\times10^{10}$ (vertical axis) in Eq.~(\ref{eq:spectrum}) as a function of $l$ (horizontal axis).}
\end{figure}

Because the induced circular polarization is found far beyond the scope of current detection, it is of no immediate importance to predict its angular power spectrum precisely.
Here, we only provide an order-of-magnitude estimate based on simple phenomenological parametrizations~\cite{Dodelson:2021}.
The power spectrum of the gravitational potential is assumed to be
\begin{align}
\frac{P_\phi(k)}{c^4} \approx \frac{9}{25} \times \frac{2\pi^2}{k^3}A_s \times T(k)^2,
\end{align}
where $A_s\approx2.1\times10^{-9}$ is the variance of primordial curvature perturbations and
\begin{align}
T(k) = \frac{\ln(2e+1.8q)}{\ln(2e+1.8q) + \left(14.2+\frac{731}{1+62.5q}\right)q^2}
\end{align}
with $q\approx(7.2~\Mpc)\times k$ is the nonoscillatory transfer function proposed in Ref.~\cite{Eisenstein:1998}.
On the other hand, the angular power spectrum of CMB temperature fluctuations is assumed to be
\begin{align}
\frac{C_l^T}{\bar{T}^2} \approx \frac{2\pi}{l(l+1)} \times \frac{A_s}{25} \times e^{-(l/l_D)^2},
\end{align}
modeling the large-scale Sachs-Wolfe plateau and the small-scale Silk damping with $l_D\approx1500$.
The resulting $c_l^\V$ is shown in Fig.~\ref{fig:spectrum} as a function of the angular scale, which reaches $l(l+1)c_l^\V/(2\pi)\approx0.55\times10^{-10}$ at $l=1000$ for example.

\section{Summary}
In summary, we showed that circular polarization of CMB is induced in principle from its temperature fluctuations when the optical Magnus effect is incorporated into weak gravitational lensing by the large-scale structure.
This is a consequence of the transverse shift of a trajectory of light depending on its helicity that requires right-handed and left-handed components at the same observation point to be sourced from different points of the surface of last scattering.
The resulting circular polarization is however suppressed by the ratio of the wavelength of light to the distance to the surface of last scattering.
A typical amplitude of the induced circular polarization relative to the mean intensity is estimated at $\sqrt{l(l+1)C_l^\V/(2\pi)}/\bar{I}\approx3.5\times10^{-35}$ for $2\pi/k=10~\mm$ and $l=1000$ as an example.
Whereas it is found far beyond the scope of current detection, our work establishes the optical Magnus effect on gravitational lensing as a new fundamental mechanism to produce circular polarization of CMB.
Furthermore, our finding applies to the gravitational wave background by regarding $\lambda\to\pm2$ as its helicity~\cite{Yoo:2012,Yamamoto:2018,Andersson:2021,Nishida:2026}, so that twice the circular polarization is induced from fluctuations of the intrinsic intensity of gravitational wave.

\begin{acknowledgments}
This work was supported by JSPS KAKENHI Grant No.~JP21K03384 and Matsuo Foundation.
\end{acknowledgments}

\end{document}